\documentclass[secnumarabic,amssymb, nobibnotes, aps, prd]{revtex4-2}

\usepackage[colorlinks=true,allcolors=blue]{hyperref}       
\usepackage{amsmath}
\usepackage{natbib}
\usepackage{adjustbox}
\usepackage{subfigure}
\usepackage{graphicx}
\usepackage{booktabs}       
\usepackage{amsfonts}       
\usepackage{nicefrac}       
\usepackage{microtype}
\usepackage{color}
\usepackage{textcomp}
\usepackage{gensymb}
\usepackage{cleveref}
\usepackage{multirow}
\usepackage{makecell}
\usepackage{array}
\newcolumntype{C}[1]{>{\centering\arraybackslash}m{#1}}

\begin{document}

\title[]{Strong Gravitational Lensing by Loop Quantum Gravity Motivated Rotating Black Holes and EHT Observations}
\author{Jitendra Kumar$^{a}$ } \email{jitendra0158@gmail.com}
\author{Shafqat Ul Islam$^{a}$ } \email{shafphy@gmail.com}
\author{Sushant~G.~Ghosh$^{a,b}$} \email{sghosh2@jmi.ac.in, sgghosh@gmail.com}

\affiliation{$^{a}$ Centre for Theoretical Physics, 
	Jamia Millia Islamia, New Delhi 110025, India}
\affiliation{$^{b}$ Astrophysics Research Centre, School of Mathematics, Statistics and Computer Science, University of KwaZulu-Natal, Private Bag 54001, Durban 4000, South Africa}	
\begin{abstract}
We investigate gravitational lensing in the strong deflection regime by loop quantum gravity (LQG)-motivated rotating black hole (LMRBH) metrics with an additional parameter $l$ besides mass $M$ and rotation $a$. The LMRBH spacetimes are regular everywhere, asymptotically encompassing the Kerr black hole as a particular case and, depending on the parameters, describe black holes with one horizon only (BH-I), black holes with an event horizon and a Cauchy horizon (BH-II), black holes with three horizons (BH-III), or black holes with no horizons (NH) spacetime. It turns out that as the LQG parameter $l$ increases, the unstable photon orbit radius $x_{ps}$, the critical impact parameter $u_{ps}$, the deflection angle $\alpha_D(\theta)$ and angular position $\theta_{\infty}$ also increases. Meanwhile, the angular separation $s$ decreases, and relative magnitude $r_{mag}$ increases with increasing $l$ for prograde motion but they show opposite behaviour for the retrograde motion. Using supermassive black holes (SMBH) Sgr A* and M87* as lenses, we compare the observable signatures of LMRBH with those of Kerr black holes. For Sgr A*, the angular position $\theta_{\infty}$ is $\in$ (16.4, 39.8) $\mu$as, while for M87* $\in$ (12.33, 29.9) $\mu$as. The angular separation $s$, for SMBHs Sgr A* and M87*, differs significantly, with values ranging $\in$ (0.008-0.376) $\mu$as for Sgr A* and $\in$ (0.006-0.282) $\mu$as for M87*. The deviations of the lensing observables $\Delta \theta_{\infty}$ and $\Delta s$ for LMRBH ($a=0.80,l=2.0$) from Kerr black holes can reach up to $10.22\mu$as and $0.241~\mu$as for Sgr A*, and $7.683~\mu$as and $0.181~\mu$as for M87*. The relative magnitude $r_{mag}$ $\in$ (0.047, 1.54). We estimate the time delay between the first and second relativistic images using twenty supermassive galactic centre black holes as lenses to find, for example, the time delay for Sgr A* and M87* can reach approximately 23.26 min and 33261.8 min, respectively. Our analysis concludes that, within the $1 \sigma$ region, a significant portion of the BH-I and BH-II parameter space agrees with the EHT results of M87* and Sgr A*. The possibility of LMRBH being a BH-III with three horizons has been almost ruled out, except for a small portion of parameter space, by $\theta_{sh}$ bounds of Sgr A* and M87*   measured by EHT. In contrast, NH without a horizon is completely ruled out. We discover that the EHT results of Sgr A* place more stringent limits on the parameter space of LMRBH black holes than those established by the EHT results of M87*. 
\end{abstract}
\maketitle

\section{Introduction}
Einstein's general relativity (GR) has been validated by the Event Horizon Telescope (EHT) observation of shadows cast by supermassive black holes M87* \cite{EventHorizonTelescope:2019dse} and SgrA* \cite{EventHorizonTelescope:2022wkp}. These also serve as further evidence of the remarkable accuracy of GR.  The black hole shadow results from strong gravitational lensing of light by the intense gravitational field of the black hole.  The extreme gravitational field of the black hole bends the light from the surrounding matter and stars, creating a dark area in the centre known as the "black hole shadow", and it is surrounded by a bright ring of light known as the "photon ring" which is caused by the bending and amplification of light around the black hole.
Utilizing gravitational lensing can offer a robust means to investigate gravitation on a large scale. By using strong-gravitational lensing by black holes and compact objects, it is possible to conduct gravity tests on a smaller scale, surpassing the limitations imposed by the Solar System \cite{EventHorizonTelescope:2019dse,EventHorizonTelescope:2022wkp}. 
In its strong field limit, gravitational lensing exhibits intriguing characteristics, making it one of the most remarkable phenomena in astronomy.
Gravitational lensing has proven to be a valuable tool for gaining insights into the structure of spacetime \cite{Einstein:1936llh,Liebes:1964zz,Mellier:1998pk,Bartelmann:1999yn,Schmidt:2008hc,2010PhRvD}. In the case of strong gravitational fields, where light passes close to the source, and the bending angle is much greater, phenomena such as shadow, photon rings, and relativistic images can occur \cite{Synge:1966okc, Gralla:2019xty, Darwin:1959, Cunha:2018acu, Bozza:2010xqn, Bozza:2001xd, Bozza:2002zj, Bozza:2002af}. 
Recent discussions have focused on strong-field gravitational lensing by black holes as another avenue to test general relativity \cite{Virbhadra:1999nm,Frittelli:1999yf, Bozza:2001xd, Bozza:2002zj, Bozza:2002af, Bozza:2010xqn, Eiroa:2002mk, Iyer:2006cn, Tsukamoto:2016jzh, Virbhadra:2007kw, Shaikh:2019jfr}. Virbhadra and Ellis \cite{Virbhadra:1999nm} provided a numerical method for studying the large deflection of light rays resulting in strong gravitational lensing. Later, Bozza  \cite{Bozza:2001xd, Bozza:2002zj, Bozza:2002af, Bozza:2010xqn} and Tsukamoto \cite{Tsukamoto:2016jzh}  analyzed the strong-field gravitational lensing analytically for general spherically symmetric and static spacetimes. Gravitational lensing by black holes has been critical in quantitative studies of the lensing by Kerr black holes \cite{Rauch:1994qd, Bozza:2008mi, Bozza:2010xqn}. With current observational facilities, the gravitational deflection of light by rotating black holes has received significant attention \cite{ Wei:2011nj, Beckwith:2004ae, Hsiao:2019ohy, Kapec:2019hro, Gralla:2019drh, James:2015yla,Cunha:2019ikd,Islam:2021ful,Ghosh:2020spb,Islam:2021dyk}. It turns out that we can understand the properties of black holes from the gravitational lensing effect. The observables in strong gravitational lensing can diagnose the properties of black holes in modified theories of gravity and compare them with their counterparts in GR.  Moreover, the photon region of gravitational lensing also provides vital properties in the black hole shadow. In addition, gravitational lensing by regular or non-singular exhibit several exciting features compared to Kerr black holes \cite{Ghosh:2020spb,Islam:2021ful,Kumar:2022fqo}.  

The prevailing belief is that singularities are an artifact of classical GR, and they can be resolved through a quantum theory of gravity \cite{Wheeler:1964}. Although a complete quantum gravity theory is not yet available, we must focus on regular models motivated by quantum arguments. Bardeen suggested the first regular solution for a black hole  \cite{Bardeen:1968}. In Bardeen's model, there are horizons, but there is no curvature singularity. Instead, the center of the black hole develops a de Sitter-like region, resulting in a black hole with a regular center. Since then, several regular black hole models have been proposed based on Bardeen's idea, which mimics the behaviour of the Schwarzschild black hole at large distances.
It is a wide belief that loop quantum gravity (LQG) could potentially address singularities in classical general relativity \cite{Ashtekar:2006wn,Ashtekar:2006es,Vandersloot:2006ws}. Because of the complexity of the complete LQG system, research has mainly focused on spherically symmetric black holes \cite{Ashtekar:2005qt,Modesto:2005zm,Boehmer:2007ket,Campiglia:2007pb,Gambini:2008dy}.
Semiclassical polymerization, which preserves the discreteness of spacetime suggested by LQG, has proved to be an effective technique for resolving the singularity issue \cite{Boehmer:2007ket,Campiglia:2007pb,Gambini:2008dy}. As different polymerizations can cause various types of regularized spacetimes, exploring a broader range of models and methods is of great interest. Building upon previous research \cite{Ashtekar:2005qt,Modesto:2005zm,Modesto:2008im,Modesto:2006mx,Boehmer:2007ket,Boehmer:2008fz,Campiglia:2007pb,Gambini:2008dy}, Peltola and Kunstatter \cite{Peltola:2009jm} used effective field theory and partially polymerized theory arguments to construct a static, spherically symmetric black hole that is asymptotically flat and encompasses the Schwarzschild black hole \cite{Peltola:2008pa,Peltola:2009jm}. Notably, unlike most regular black holes with two horizons, this LQG-corrected black hole has just one horizon.  However, astrophysical observations cannot test non-rotating black holes, as the black hole spin plays a critical role in any astrophysical process. Using the modified Newman-Janis algorithm, prompted us to generalise these regular solutions to the axially symmetric case or the Kerr-like solution -- LQG-motivated rotating black holes (LMRBH) \cite{Kumar:2022vfg,Islam:2022wck}. Testing LMRBH metrics with astrophysical observations, like the EHT observations, is crucial. The spin is essential as it signifies the current-dipole moment of the gravitational field produced by a compact object. It serves as the primary correction to the term mass-monopole.
\begin{figure*}[t]
	\begin{center}
		    \includegraphics[scale=0.99]{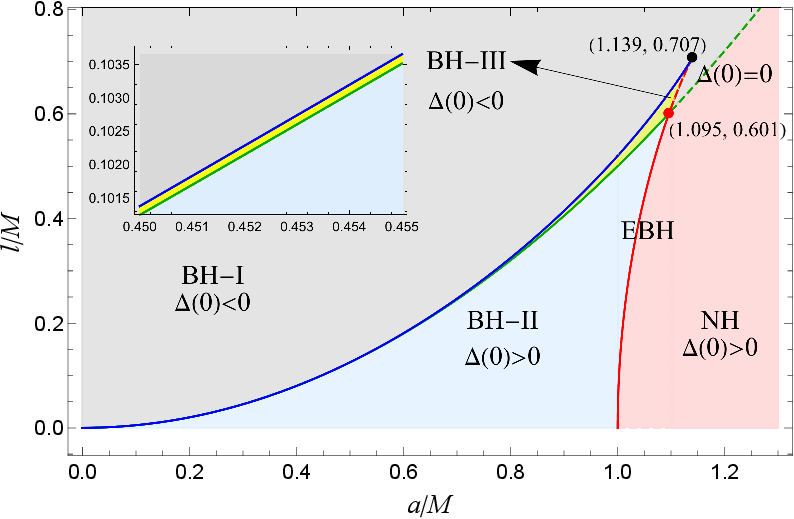}
	\end{center}
	\caption{Parameter space ($a/M,l/M$) for LMRBH spacetime \cite{Kumar:2022vfg}. The red line corresponds to the extremal black holes with degenerate horizons, where the outer two horizons merge. The blue line also corresponds to the black hole with degenerate horizons, but unlike the red line, the inner two horizons merge. For the green line, $\Delta(r)=0$ admits two positive and the third root at $r=0$, while for the dashed green line, we have $r=0$ as the only root. The black dot corresponds to the black hole with three degenerate horizons located at $r=0.5 M$.}\label{parameter}		
\end{figure*}  

The prime aim of this investigation is to explore the gravitational lensing properties of recently obtained LMRBH metric \cite{Kumar:2022vfg,Islam:2022wck} and compare them to the gravitational lensing by Kerr black holes. In addition, we investigate the observable characteristics of LMRBH versus Kerr black holes when supermassive black holes such as Sgr A* and M87* act as lenses. Notably, although strong deflection lensing effects by LMRBH black holes could be detected using the Event Horizon Telescope (EHT), distinguishing between two black holes is challenging because of deviations being on the order of $\mathcal{O}(\mu$as). 

The paper is organized as follows: We briefly review the horizon structure and calculate the deflection angle by LMRBH spacetime in Sect.~\ref{Sec2}.  The strong-lensing observables by the LMRBH, including the image positions $\theta_{\infty}$, separation $s$, and magnifications $\mu_n$  are also part of Sect.~\ref{Sec2}.  The time delay between the first and second images on the same side of the source has  been calculated for supermassive black holes SgrA*, M87* and  those at the centers of 19 other galaxies in Sect.~\ref{Sec2}.  A numerical analysis of  the observables by  taking the supermassive black holes  Sgr A* and M87* as the lens is part of  Sect.~\ref{Sec3}. The constraints on the LMRBH parameters inferred using black hole shadow observational data of Sgr A* and M87* are discussed in  Sect.~\ref{Sec4}. Finally, we summarize our results to end the paper in Sect.~\ref{Sec5}. 

 Throughout this paper, unless otherwise stated, we adopt natural units ($8 \pi G\; =\; c\;=\; 1$)
\section{Gravitational Lensing by LQG-motivated rotating black holes}\label{Sec2}
We derived the metric for the LQG-motivated rotating black hole (LMRBH) by building upon a partially polymerized static and spherically symmetric black hole solution \cite{Peltola:2009jm, KumarWalia:2022ddq}. To construct the rotating spacetime LMRBH, we employed the revised Newman–Janis algorithm (NJA) described in \cite{Azreg-Ainou:2014pra, Azreg-Ainou:2014aqa}. This procedure has been successful in generating imperfect fluid rotating solutions in Boyer-Lindquist coordinates from spherically symmetric static solutions, and it can also produce generic rotating regular black hole solutions \cite{Azreg-Ainou:2014pra, Azreg-Ainou:2014aqa, Ghosh:2021clx, Ghosh:2014pba, Mazza:2021rgq}.  We express the metric of the LMRBH in the Boyer-Lindquist form as  \cite{Kumar:2022vfg, Islam:2022wck}.
\begin{eqnarray}\label{metric3}
ds^2 &=& -\left[1-\frac{2M(r)\sqrt{r^2+l^2}}{\rho^2}\right] dt^2+ \frac{\rho^2}{\Delta} dr^2 +\rho^2 d\theta^2  \nonumber\\ && - \frac{4aM(r) \sqrt{r^2+l^2}}{\rho^2} \sin^2\theta dtd\phi+ \frac{\mathcal{A}\sin^2\theta~}{\rho^2} d\phi^2
\end{eqnarray}
where
\begin{eqnarray}
M(r) &=& M-\frac{r-\sqrt{r^2+l^2}}{2} ~~~~~
\rho^2 = r^2+l^2 +a^2\cos^2\theta,\nonumber \\
\Delta &=& r^2+l^2 +a^2 -2 M(r) \sqrt{r^2+l^2},\nonumber \\
\mathcal{A} &=& (r^2+l^2 +a^2)^2-a^2 \Delta \sin^2\theta.
\end{eqnarray} 

The rotating metric, derived from the NJA, captures important elements of LQG, such as a transition surface at the black hole centre and the global regularity of spacetime.  It is noteworthy that LMRBH  (Eq.~(\ref{metric3})) encompasses the Kerr spacetime \cite{Kerr:1963ud} in the limit $l \to 0$ and spherical LQG black hole \cite{Peltola:2009jm,KumarWalia:2022ddq} in the limit $a \to 0$. When $a=M= l= 0$, then Eq.~(\ref{metric3}) gives flat spacetime.
\begin{figure*}[t] 
	\begin{centering}
		\begin{tabular}{cc}
		    \includegraphics[scale=0.75]{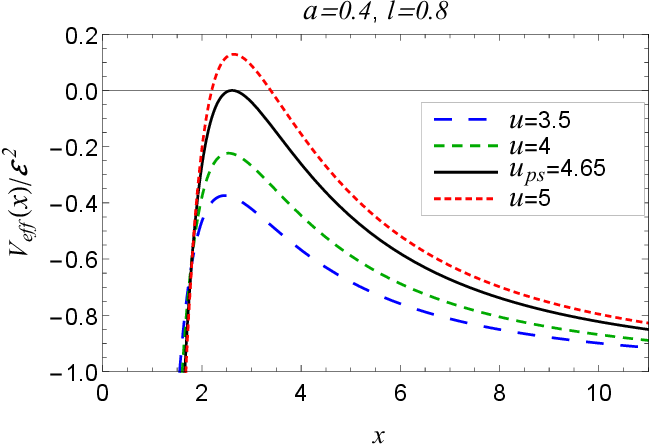}\hspace{0.5cm}
		    \includegraphics[scale=0.75]{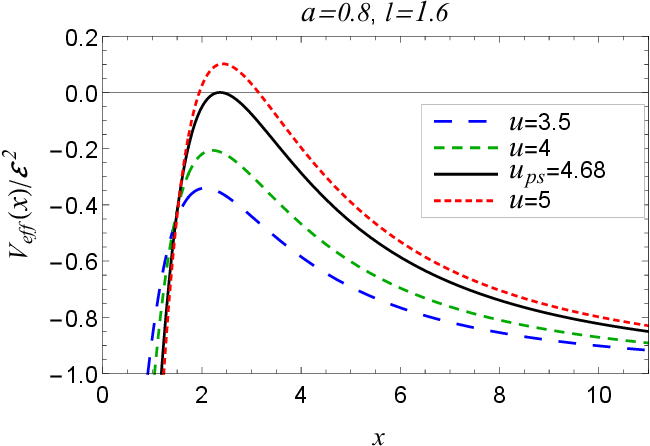}
			\end{tabular}
	\end{centering}
	\caption{Effective potential for photons, in LMRBH spacetime, having different impact parameters $u$. Black curves correspond to the photons whose impact parameter $u$ is exactly equal to the critical impact parameter $u_{ps}$. These photons revolve around the black hole in unstable photon circular orbits at radial distance $x=x_{ps}$. Photons with $u \approx u_{ps}$ but $u>u_{ps}$ make several loops around the black hole and are scattered to infinity after reaching some closest approach distance $x_0$. These photons form the (strong) gravitationally lensed image of the source. Photons with $u<u_{ps}$ fall into the black hole. }\label{plot1}
\end{figure*}
The horizons of the LMRBH are determined by the roots of the null surface $\Delta(r)=0$, which is a coordinate singularity of Eq.(\ref{metric3}). Depending on the values of $a$ and $l$, up to three real roots may exist, with one to three positive roots. Only positive roots correspond to horizons, and we label them as $r_1$, $r_2$, and $r_3$ with $r_3 \le r_2 \le r_1$, where $r_1$ is the event horizon and $r_2$, if present, is the Cauchy horizon. The additional root $r_3$ lies inside the Cauchy horizon. The parameter space ($a,l$) for the LMRBH is illustrated in Fig.~\ref{parameter}. There exists a critical value of $l$ ($a$) on the red line, denoted by $l_c$ ($a_c$), such that $\Delta(r)=0$ has a double root, for given $a$ ($l$), corresponding to an extremal LMRBH with degenerate horizons. For $a<a_c$ ($l>l_c$), $\Delta(r)=0$ has two simple positive roots, corresponding to LMRBHs with Cauchy and event horizons (BH-II). In contrast, for $a>a_c (l<l_c)$, $\Delta(r)=0$ has no positive roots, corresponding to no-horizon (NH) spacetimes. Similarly, for a given $a$ ($l$), we can find the critical value of $l$ ($a$) on other transition lines and dots in Fig.~\ref{parameter}. We consider four regions for our study, namely black holes with only one horizon (BH-I), black holes with an event horizon and Cauchy horizon (BH-II), black holes with three horizons (BH-III), and black holes with no horizon (NH). The coloured lines in Fig.~\ref{parameter} denote the boundaries that divide these regions (see Kumar et al. (\cite{Kumar:2022vfg}) for more information).
\begin{figure*} 
	\begin{centering}
		\begin{tabular}{p{9cm} p{9cm}}
		    \includegraphics[scale=0.7]{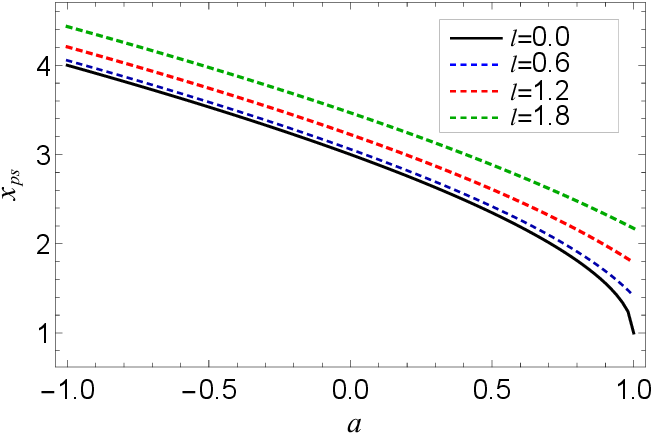}&
		    \includegraphics[scale=0.7]{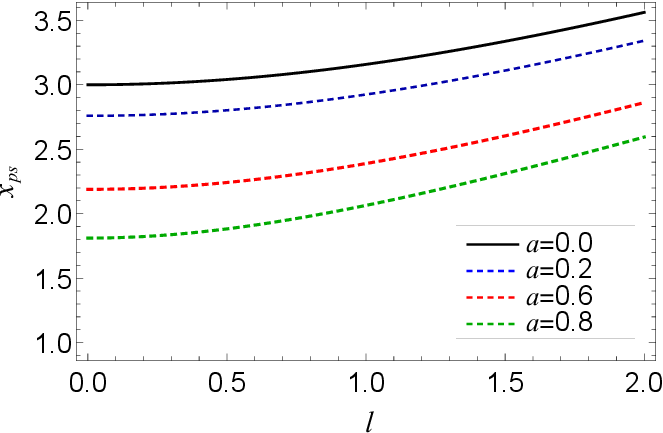}
			\end{tabular}
	\end{centering}
	\caption{Behavior of the photon sphere radius $x_{\text{ps}}$ with respect to the parameter $a$ (left) for different $l$ and with respect to the parameter $l$ (right) for different $a$ for LMRBH spacetime. Negative values of $a$ in the left diagram correspond to the retrograde motion of the photon.}\label{plot4}
\end{figure*}
\begin{figure*}
	\begin{centering}
		\begin{tabular}{p{9cm} p{9cm}}
		    \includegraphics[scale=0.7]{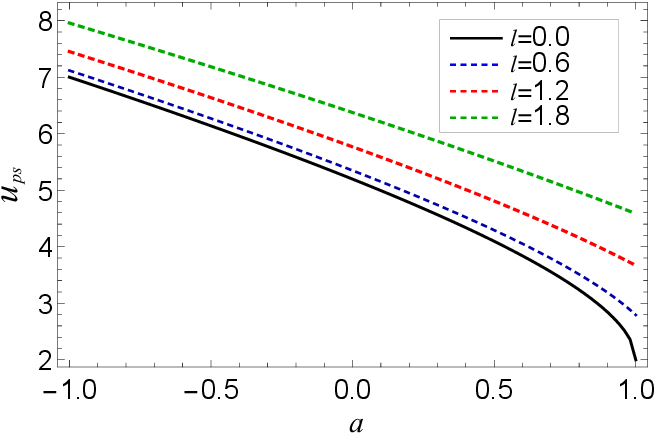}&
		    \includegraphics[scale=0.7]{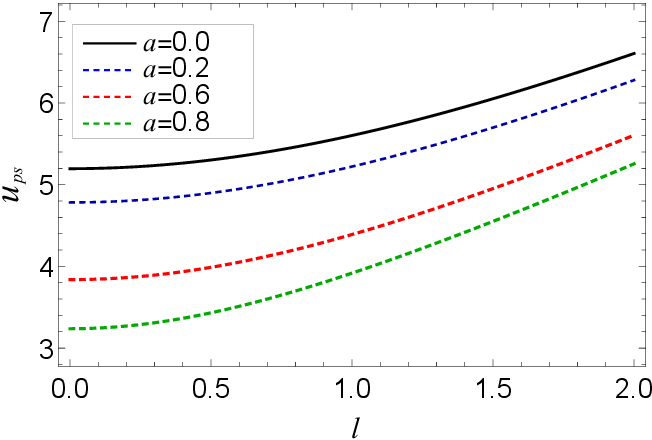}
			\end{tabular}
	\end{centering}
	\caption{Behavior of the critical impact parameter $u_{\text{ps}}$ with respect to the parameter $a$ (left) for different $l$ and with respect to the parameter $l$ (right) for different $a$ for LMRBH spacetime. Negative values of $a$ in the left diagram correspond to the retrograde motion of the photon.}\label{plot5}
\end{figure*}
The presence of a black hole has an influential impact on the motion of nearby photons. The null geodesics that describe photon orbits around black holes are crucial for observing gravitational effects caused by the black hole. Two linearly independent killing vectors, $\eta^{\mu}_{(t)}=\delta^{\mu}_t $ and $\eta^{\mu}_{(\phi)}=\delta^{\mu}_{\phi}$, associated with the time translation and rotational invariance,  are admitted by the black hole metric (\ref{metric3}) \cite{Chandrasekhar:1985kt}. The total energy $\mathcal{E}$ and the angular momentum $\mathcal{L}$, two conserved quantities corresponding to the Killing vectors, govern the photon's path. We identify $u = \mathcal{L}/\mathcal{E}$ as the impact parameter and employ the Hamilton-Jacobi approach to establish the relationship between the effective potential $V_{\text{eff}}$ and impact parameter $u$. The relationship between $u$ and $V_{\text{eff}}$ determines the orbit's qualitative characteristics. Photons with different impact parameters approach the black hole and are deflected by its strong gravitational field, reaching a minimum distance of $r_0$. It is critical to understand that light rays with impact parameters smaller than the critical value $u_{ps}$ fall into the event horizon, leading to a dark spot on the observer's sky. On the other hand, light rays with impact parameters slightly larger than the critical value make several
loops around the black hole and are scattered to reach the observer at infinity (cf. Fig. \ref{plot1}). Finally, photons with impact parameters equal to the critical value follow unstable circular photon orbits with a constant radius $r_{ps}$ (cf. Fig. \ref{plot1}). By considering light rays near the critical impact parameter, we investigate the effect of the parameter $l$ on strong gravitational lensing using the LMRBH model to represent the black holes. The LMRBH metric, similar to the Kerr metric, exhibits reflection symmetry $\theta \to \pi - \theta$, allowing for studying light ray motion in the equatorial plane where they initially reside. To explore the strong gravitational lensing effects, we follow the approach in Refs. \cite{Islam:2021ful, Islam:2021dyk, Bozza:2002zj}. 

We  measure  the quantities $r, a, l$,  and $t$ in units of  $M$ \cite{Bozza:2002zj} such that 
\[
r/M \to x, ~~~~a/M \to a, ~~~~l/M \to l \;\;\; and ~~~t/M \to t, 
\]
and use $x$ instead of radius $r$  to rewrite the LMRBH metric (\ref{metric3}) in the equatorial plane ($\theta= \pi/2$) as 
\begin{eqnarray}\label{NSR}
\mathrm{ds^2}=-A(x)dt^2+B(x) dx^2 +C(x)d\phi^2-D(x)dt\,d\phi,
\end{eqnarray}
where
\begin{eqnarray}\label{compo}
A(x) &=& 1-\frac{2M(x)\sqrt{x^2+l^2}}{\rho^2},~~~~
B(x)=\frac{{\rho}^2}{\Delta}, \nonumber\\
C(x) &=& \frac{\mathcal{A}}{{\rho}^2},~~~~
D(x) = 4aM(x) \sqrt{x^2+l^2},    
\end{eqnarray}
and ${\rho}^2 = x^2+l^2 $ and $\mathcal{A} =(x^2+l^2 +a^2)^2-a^2\Delta $.  

\begin{figure*} 
	\begin{centering}
		\begin{tabular}{p{9cm} p{9cm}}
		    \includegraphics[scale=0.75]{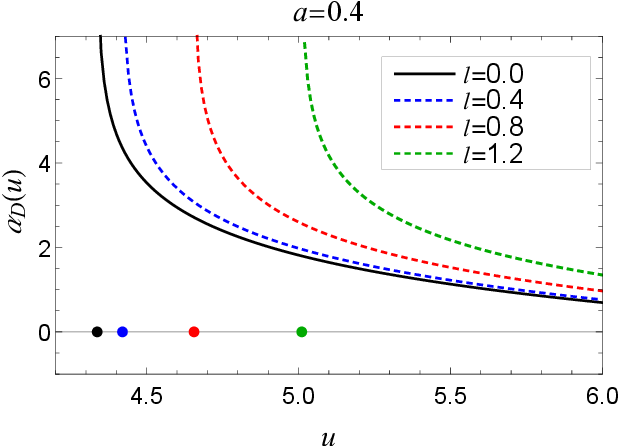}&
		    \includegraphics[scale=0.75]{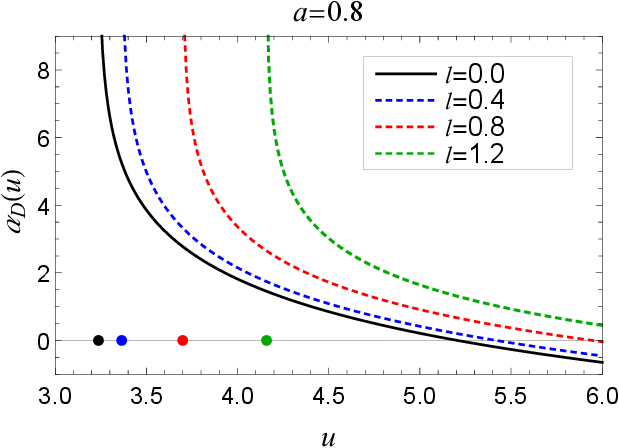}
			\end{tabular}
	\end{centering}
	\caption{ Variation of deflection angle, in strong field limit, for LMRBH spacetime as a function of the impact parameter $u$ for different values of the parameters $a$ and $l$. Dots on the horizontal axis represent the values of the critical impact parameter $u_{\text{ps}}$ at which the deflection angle diverges.}\label{plot6}
\end{figure*}	
A photon from a source travels towards a black hole until it is at least at a distance of $x_0$ away from it before being redirected by the black hole's gravitational field to reach the observer at infinity. The impact parameter, which is the perpendicular distance from the black hole's centre of mass to the initial direction of the photon at infinity, stays constant throughout the trajectory because the approach phase resembles the departure phase with time-reversed due to spacetime symmetry. Furthermore, the minimum approach distance $x_0$ marks the turning point and hence vanishing  effective potential, which gives an expression for the impact parameter $u$ in terms of the closest approach distance $x_0$ as
\begin{eqnarray}
u &=& \frac{\mathcal{L}}{\mathcal{E}} = \frac {1}{(x_{0} - 2)\sqrt {l^2 + x_{0}^2}} \nonumber \\ 
&&\Bigg[\left (l^2 + x_{0}^2 \right)\sqrt {a^2 + (x_{0} - 2)\sqrt {l^2 + x_{0}^2}} \nonumber\\ &+& 
   a\left (x_{0}\sqrt {l^2 + x_{0}^2} - 
      2\sqrt {l^2 + x_{0}^2} - 
      l^2 - x_{0}^2 \right) \Bigg]. 
\end{eqnarray}
By restricting the light rays to the equatorial plane, the unstable circular photon orbits radius $x_{ps}$ in terms of metric components, given by Eq.~(\ref{compo}), is the solution  of the  equation \cite{Bozza:2002af}  
\begin{eqnarray}\label{ps}
A(x)C'(x)-A'(x)C(x) + u(A'(x)D(x) - A(x)D'(x)) &=& 0,
\end{eqnarray}
which implies 
\begin{eqnarray}\label{ps1}
\frac{1}{(x - 2)\left (l^2 + 
     x^2 \right)} \Bigg[2 a\left (l^2 + 
       2 x \right)\sqrt {a^2 + (x - 2)\sqrt {l^2 + x^2}} \nonumber\\ - 
   2 a^2\left (l^2  + 
      2 x \right) - (x - 2)\left (l^2 - 
       2 (x - 3) x \right)\sqrt {l^2 + x^2}\Bigg] &=& 0
\end{eqnarray}

The unstable photon orbit radius is the largest root of the Eq.~\ref{ps1}. It turns out that the photon sphere depends on the parameter $l$ and the rotation parameter. By fixing the winding of light rays to be counterclockwise, we assign  $a>0$ (prograde orbits), if the black hole also rotates in the counterclockwise direction and  $a<0$ (retrograde) if the black rotates in the opposite direction of photon winding i.e., clockwise. Fig.~\ref{plot4} shows the decrease in radius $x_{ps}$ with the rotation parameter $a$ while the opposite behaviour concerning the parameter $l$ and suggests that the $x_{ps}$ of LMRBH is greater than the Kerr black holes. Fig.~\ref{plot4} also suggests that the photons forming prograde orbits can get closer to the black hole than the photons forming retrograde orbits. The critical impact parameter $u_{ps}$ is the parameter for which the closest approach distance $x_0$ equals the photon orbit radius $x_{ps}$. We have plotted the critical impact parameter (cf. Fig.~\ref{plot5}) with varying $a$ and $l$ and found that it varies similarly to unstable photon orbit radius.

Moreover, the deflection angle of a photon moving in the equatorial plane of the LMRBH spacetime can be obtained by using the null geodesic equations, which are first-order ordinary differential equations \cite{Chandrasekhar:1985kt}. The light bending angle in a general rotating stationary spacetime described by the line element (\ref{NSR}), for a closest distance approach $x_0$ is given by \cite{Islam:2021ful, Islam:2021dyk, Bozza:2002zj, Virbhadra:1998dy}
\begin{eqnarray}\label{bending2}
\alpha_{D}(x_0) &=& -\pi + 2\int_{x_0}^{\infty} dx 
\nonumber\\ && \frac{\sqrt{A_0 B }\left(2Au+ D\right)}{\sqrt{4AC+D^2}\sqrt{A_0 C-A C_0+u\left(AD_0-A_0D\right)}},
\nonumber\\
\end{eqnarray}
This integral is non-trivial to solve, and hence, we expand it near the unstable photon sphere radius \cite{Virbhadra:1999nm,Claudel:2000yi,Bozza:2002zj} by defining a new variable $z=1-x_0/x$ in strong deflection limit (SDL)  \cite{Tsukamoto:2016jzh,Zhang:2017vap}. This technique  not only shows the behaviour of photons near the photon sphere but also provides an analytical representation of the deflection angle as \cite{Bozza:2002zj,Kumar:2020sag,Islam:2020xmy}
\begin{eqnarray}\label{def4}
\alpha_{D}(u) &=& \bar{a} \log\left(\frac{u}{u_{ps}} -1\right) + \bar{b} + \mathcal{O}(u-u_{ps}),  
\end{eqnarray}  

where $\bar{a}$, $\bar{b}$ are the lensing coefficients. The details of this calculation can be found in \cite{Bozza:2002zj,Kumar:2020sag,Islam:2020xmy}. The deflection angle increases as $x_0$ approaches $x_{ps}$, eventually exceeding $2\pi$ radians and diverging logarithmically at $x_0 = x_{ps}$.  We can investigate the deflection angle of strong gravitational lensing by LMRBH and compare it with the analogous results of the Kerr black hole. Like Kerr black hole, the deflection angle for LMRBH spacetime in SDL (cf.~Fig.~\ref{plot6}) is more than $2\pi$. The effect of the LMRBH parameter $l$ and spin $a$ can be seen in Fig. \ref{plot6}. The deflection angle diverges at larger $u_{ps}$ for larger $l$ while at smaller $u_{ps}$ for larger $a$. 
\begin{figure*}
\begin{center}
\includegraphics[scale=0.8]{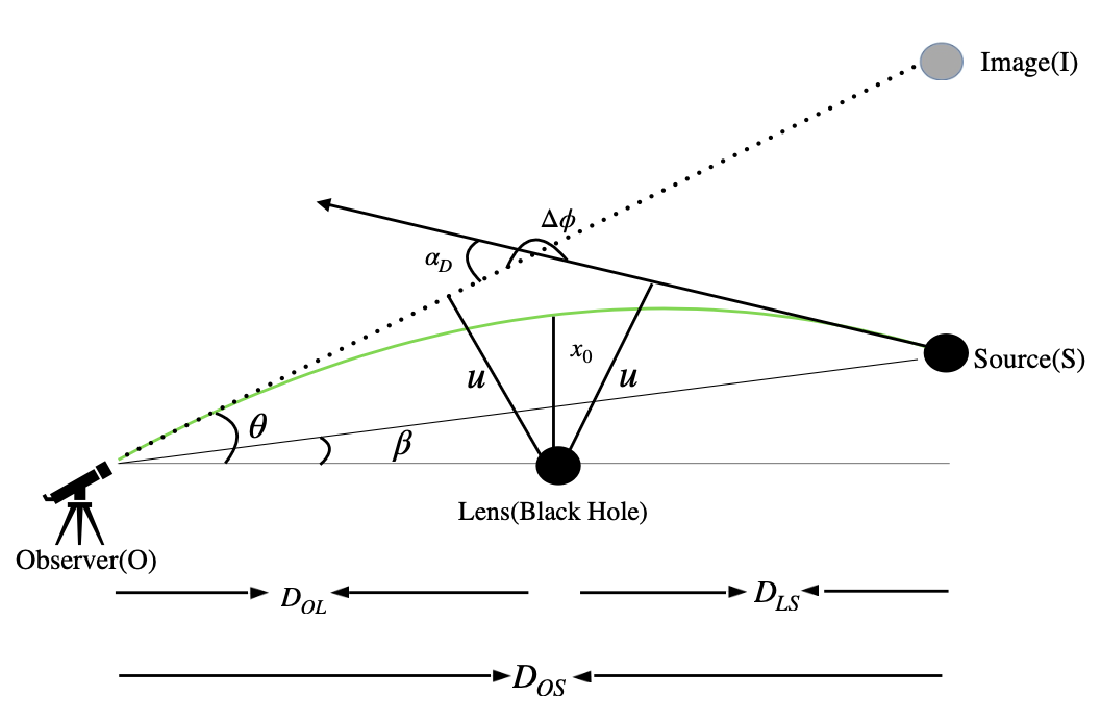}
\end{center}
\caption{Formation of primary images of source space (S) in case of gravitational lensing. Light rays are deviated by the black hole to be observed at an angular position $\theta$ by the observer space (O).} \label{plot5b}  
\end{figure*}
As can be seen in Eq.~(\ref{def4}) if a light beam's impact parameter is sufficiently close to its critical value, it may also approach the unstable photon orbit radius and make one, two, or even more turns around the lens before it reaches the observer. As a result, a strong gravitational field can generate many images, creating what may appear to be an endless series of images. The two infinite sets of relativistic pictures represent the clockwise and anticlockwise winds of the black hole. In Fig. \ref{plot6}, we show that for higher impact parameter values, the deflection angle for some parameters can become negative, which shows that the path of the light or particles is bent opposite the black hole's centre, causing photons to be deflected away from the black hole's location. But for our configuration in which the black hole lies between a light source and observer, we calculated the deflection angle using the expression in Eq.~(\ref{def4}), which is valid for SDL. However, this expression of the deflection angle in SDL technique is valid only in close vicinity to the critical impact parameter. In reality, a negative deflection angle value is not achievable. Still, as is usually done in the literature, one can utilise the weak deflection limit technique to get the deflection angle at larger impact parameters  \cite{Bozza:2010xqn}.

\subsection{Lens equation and observables}
The lens geometry is essential in understanding black hole lensing because it can be used to identify the exact positions and magnifications of relativistic images in particular. Since the light source and observer are placed far enough from the black hole for the gravitational forces around them to be insufficient, the lens configuration we are interested in positions the black hole between a light source and observer \cite{Bozza:2002zj,Bozza:2008ev} (cf. Fig.~\ref{plot5b}). Assuming that the source and observer are almost aligned, the lens equation reads
\begin{eqnarray}\label{lenseq}
\beta &=& \theta -\frac{D_{LS}}{D_{OL}+D_{LS}} \Delta\alpha _n,
\end{eqnarray}
\begin{figure*}
	\begin{centering}
		\begin{tabular}{p{9cm} p{9cm}}
		    \includegraphics[scale=0.60]{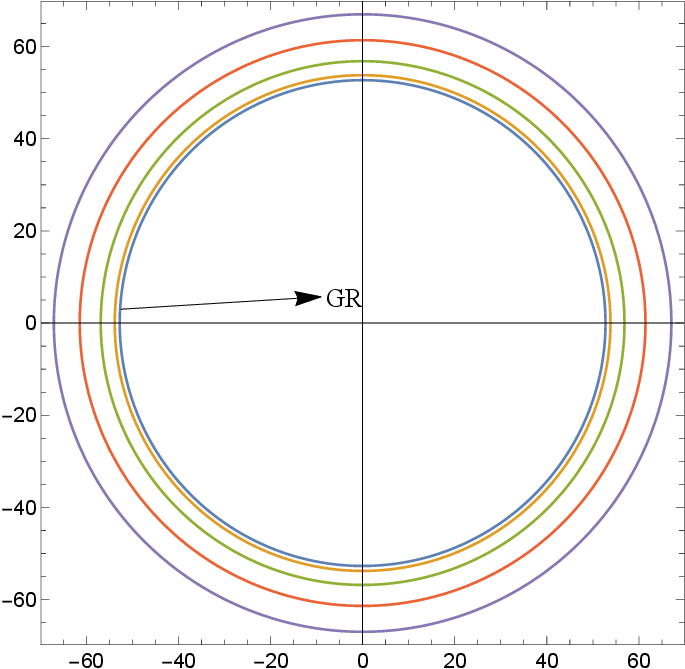}&
			\includegraphics[scale=0.60]{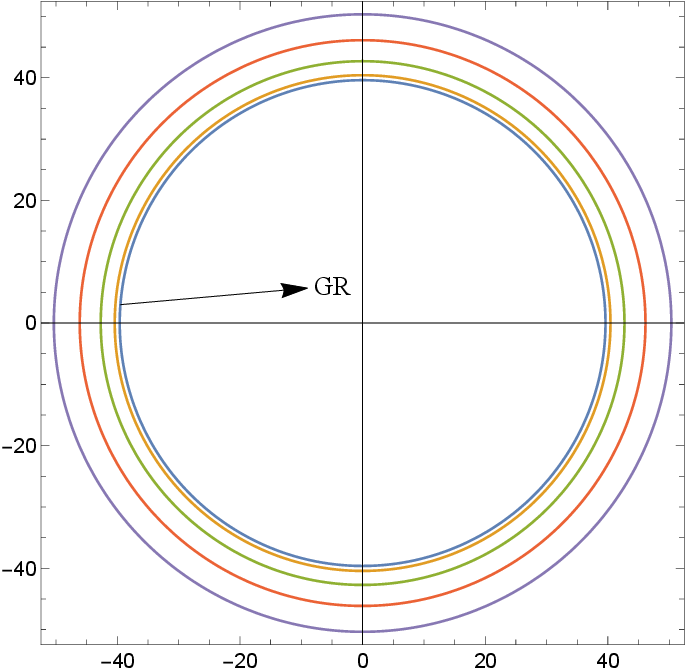}			
		\end{tabular}
	\end{centering}
	\caption{Formation of outermost relativistic Einstein ring  for spherically symmetric case and  values of  $l$ taken to be $0,~0.5,~1.0,~1.5,~2.0$ (innermost to outermost) for Sgr A* (\textit{left}) and  M87* (\textit{right}). The inner rings, respectively, correspond to the case when Sgr A* and M87* are considered as Schwarzschild black holes ($a=l=0$).}\label{ERing}		
\end{figure*} 
Instead of using a full deflection angle, we used an offset of deflection angle $\Delta\alpha_{n}=\alpha-2n\pi$ with $n$ being an integer with $n \in N$, and $0 < \Delta\alpha_n \ll 1$. Here, $\beta$ and $\theta$ are the angular positions of the source and image from the optical axis, respectively. The distances of the source and lens from the observer are given by $D_{OS}$ and $D_{OL}$, respectively (cf. Fig.~\ref{plot5b}). 

Next, we shall estimate the observables for the strong gravitational lensing by LMRBH spacetime as in \cite{Islam:2021dyk,Ghosh:2020spb,Bozza:2002af,Bozza:2002zj}. Using the lens Eq.~(\ref{lenseq}) and Eq.~(\ref{def4}), and following the condition where the source, lens and the observer are aligned, the angular separation between the lens and the $n^{th}$ image is given by~\cite{Bozza:2002af}  
\begin{eqnarray}\label{angpos}
\theta_n &=& \theta_n{^0} + \Delta\theta_n,
\end{eqnarray}
where 
\begin{eqnarray}
\theta_n{^0} &=& \frac{u_m}{D_{OL}}(1+e_n), \label{postion}\\
\Delta\theta_n &=& \frac{D_{OL}+D_{LS}}{D_{LS}}\frac{u_me_n}{\bar{a} D_{OL}}(\beta-\theta_n{^0}),\\
e_n &=& \text{exp}\left({\frac{\bar{b}}{\bar{a}}-\frac{2n\pi}{\bar{a}}}\right).
\end{eqnarray}
Here $\theta_n{^0}$ is the angular position of the image when a photon
encircles complete $2n\pi$ and the second term in Eq.~(\ref{angpos}) is the extra term exceeding $2n\pi$ such that $\theta_n{^0} \gg \Delta\theta_n $ \cite{Bozza:2002zj}. When $\beta = 0$ is entered into Eq.~(\ref{angpos}), one can obtain the angular radius of the Einstein rings for spherically symmetric cases \cite{Einstein:1936llh,Liebes:1964zz,Mellier:1998pk,Bartelmann:1999yn,Schmidt:2008hc,2010PhRvD}. Einstein rings for Sgr A* and M87* are plotted in Fig.~\ref{ERing}. 
\begin{figure*}
	\begin{centering}
		\begin{tabular}{p{9cm} p{9cm}}
		    \includegraphics[scale=0.75]{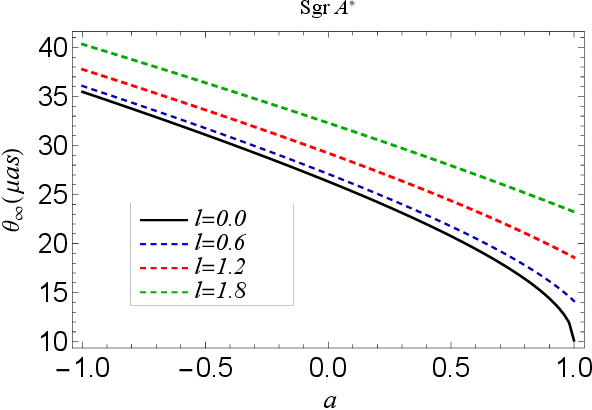}&
		    \includegraphics[scale=0.75]{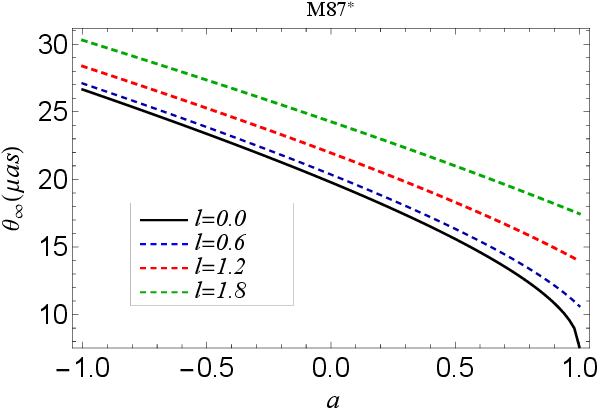}\\
		    \includegraphics[scale=0.75]{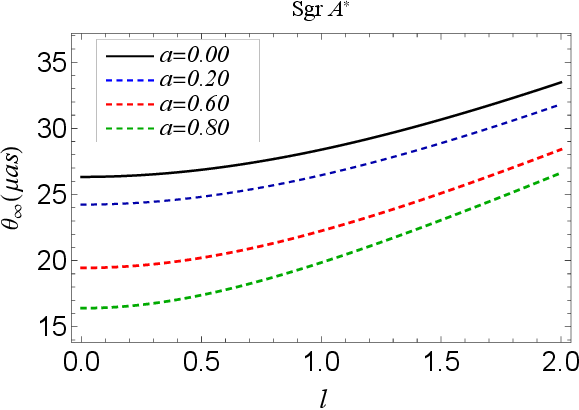}&
		    \includegraphics[scale=0.75]{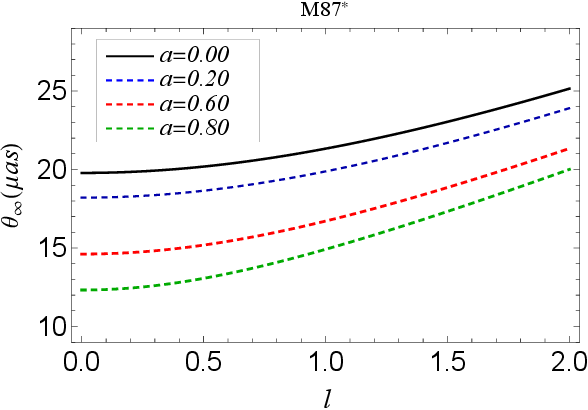}\\
			\end{tabular}
	\end{centering}
	\caption{Behavior of strong lensing observables $\theta_{\infty}$ in strong field limit, as a function of the parameters $a$ and $l$ by considering that the spacetime around the compact objects at the centers of Sgr A*(left panel) and M87*(right panel) is LMRBH spacetime.}\label{plot7}
\end{figure*} 
\begin{figure*}
	\begin{centering}
		\begin{tabular}{p{9cm} p{9cm}}
			\includegraphics[scale=0.75]{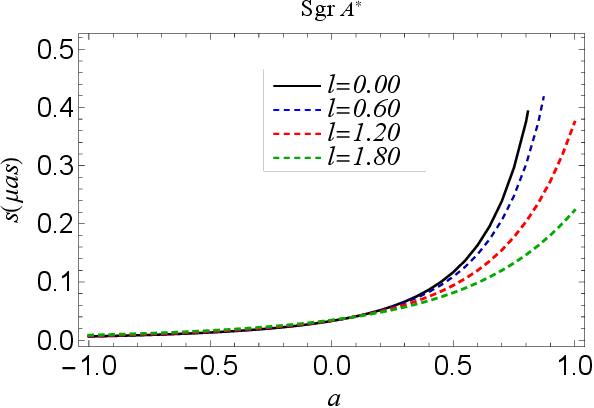}&
			\includegraphics[scale=0.75]{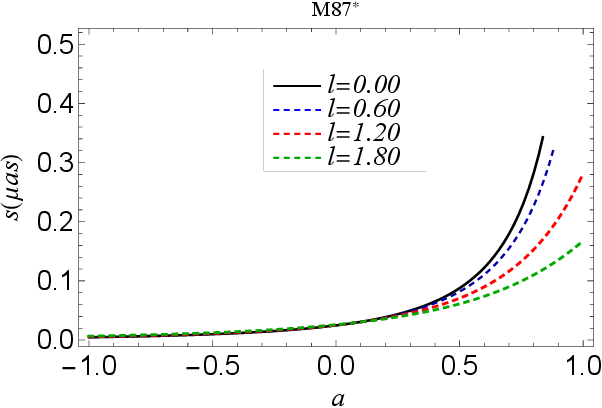}\\
			\includegraphics[scale=0.75]{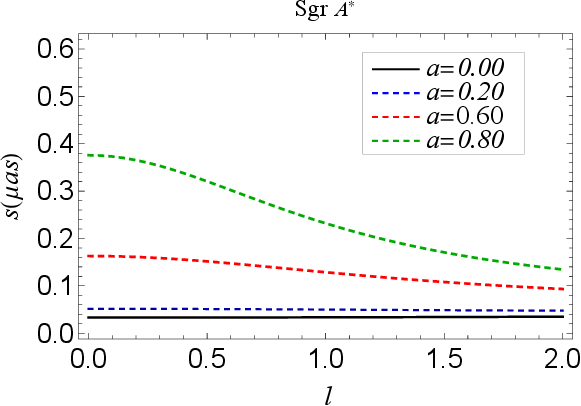}&
			\includegraphics[scale=0.75]{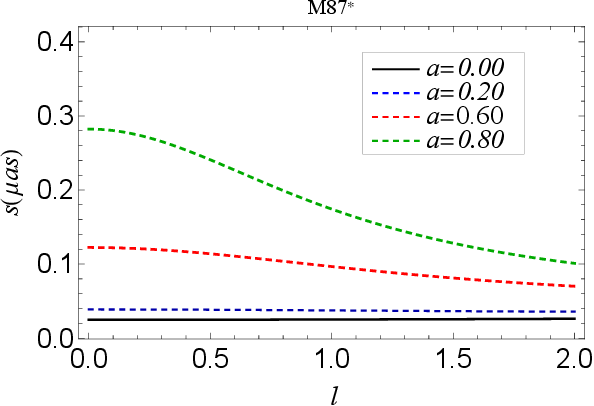}
			\end{tabular}
	\end{centering}
	\caption{Behavior of strong lensing observables  $s$ in strong field limit, as a function of the parameters $a$ and $l$ by considering that the spacetime around the compact objects at the centers of Sgr A*(left panel) and M87*(right panel) is LMRBH spacetime.}\label{plot7b}
\end{figure*}    

\begin{table*}[htb!]
\resizebox{18cm}{!}{
 \begin{centering}	
	\begin{tabular}{p{2cm} p{2cm} p{2cm} p{2cm} p{2cm} p{2cm} p{2cm}}
\hline\hline
\multicolumn{2}{c}{}&
\multicolumn{2}{c}{Sgr A*}&
\multicolumn{2}{c}{M87*}\\
{$a$ } & {$l$}& {$\theta_{\infty}$($\mu$as)} & {$s$ ($\mu$as)} & {$\theta_{\infty}$($\mu$as)}  & {$s$ ($\mu$as) } & {$r_{mag}$} \\ \hline
\hline
\multirow{7}{*}{-0.8}&0.0& 33.760 & 0.008 & 25.365 & 0.006 & 8.522 \\
& 0.5 & 34.194 & 0.009 & 25.690 & 0.006 & 8.492 \\ 
& 1.0 & 35.440 & 0.009 & 26.627 & 0.007 & 8.415 \\ 
& 1.5 & 37.363 & 0.010 & 28.071 & 0.008 & 8.312 \\ 
& 2.0 & 39.804 & 0.012 & 29.906 & 0.009 & 8.203 \\             
\hline 
\multirow{7}{*}{-0.4}&0.0& 30.188 & 0.016 & 22.681 & 0.012 & 7.755 \\
& 0.5 & 30.664 & 0.016 & 23.038 & 0.012 & 7.736\\
& 1.0 & 32.021 & 0.017 & 24.058 & 0.013 & 7.687 \\ 
& 1.5 & 34.092 &  0.018 & 25.614 & 0.014 & 7.626\\ 
& 2.0 & 36.691 & 0.020 & 27.567 & 0.015 & 7.568 \\     
\hline
\multirow{7}{*}{0.0}&0.0& 26.330 & 0.033 & 19.782 &0.025& 6.822\\ 
& 0.5 & 26.869 & 0.033 & 20.187 & 0.025 & 6.822\\
& 1.0 & 28.389& 0.033 & 21.330 & 0.025 & 6.826\\
& 1.5 & 30.668 & 0.034 & 23.041 & 0.026 & 6.836\\
& 2 & 33.475 & 0.035 &  25.151 & 0.026 & 6.854\\ 
\hline
\multirow{7}{*}{0.4}& 0.0 & 21.977 & 0.087 & 16.511& 0.065 & 5.587\\
& 0.5 &  22.631 & 0.084 & 17.002 & 0.063 & 5.634 \\
& 1.0 & 24.427 & 0.078 & 18.352 & 0.058 & 5.752\\ 
& 1.5 & 27.027 & 0.071 & 20.306 & 0.053 & 5.897 \\ 
& 2 & 30.131 & 0.065 & 22.638 & 0.049 & 6.04 \\   
\hline
\multirow{7}{*}{0.8}& 0.0 & 16.404 & 0.376 & 12.325 & 0.282 & 3.561 \\ 
& 0.5 & 17.392 & 0.321 & 13.067 &  0.241 & 3.801\\ 
& 1.0 & 19.853 & 0.232 & 14.916 & 0.174 & 4.279 \\ 
& 1.5 & 23.062 & 0.171 & 17.327 & 0.128 & 4.732 \\ 
& 2 & 26.631 & 0.134 & 20.008 & 0.101 & 5.099 \\                       
		\hline\hline
	\end{tabular}
\end{centering}
}	
	\caption{Estimates for the lensing observables by considering supermassive black holes at the center of nearby galaxies as LMRBHs. We  measure  the quantities $a$ and $l$ in units of the mass of the black hole, $M$.  }\label{table1}
\end{table*}

\begin{table*}[htb!]
\resizebox{18cm}{!}{
 \begin{centering}	
	\begin{tabular}{p{1.5cm} p{1.5cm} p{2.0cm} p{2.0cm} p{2.0cm} p{2.0cm} p{2.0cm} p{2.0cm} p{2.0cm} p{2.0cm} p{2.0cm}}
\hline\hline
\multicolumn{2}{c}{}&
\multicolumn{2}{c}{Sgr A*}&
\multicolumn{2}{c}{M87*}\\
{$a$ } & {$l$}& 
{$\Delta\theta_{\infty}$($\mu$as)} & {$\Delta s$ ($\mu$as)} & {$\Delta\theta_{\infty}$($\mu$as)}  & {$\Delta s$ ($\mu$as) } &  
{$\Delta r_{mag}$} \\ \hline
\hline
\multirow{7}{*}{-0.40} 
& 0.50  &  0.476 & 0.0003 & 0.358 & 0.0002 & -0.019\\                 
& 1.00  & 1.833 & 0.001 & 1.377 & 0.0009 & -0.068  \\              
& 1.50  & 3.910 & 0.003 & 2.933 & 0.002 & -0.130 \\
& 2.00  &  6.503 & 0.004 & 4.886 & 0.003 & -0.188 \\
\hline        
\multirow{7}{*}{0.40} 
& 0.50 & 0.653 & -0.002 & 0.491 & -0.002 & 0.047 \\                         
& 1.00  & 2.451 & -0.009 & 1.841 & -0.007 & 0.165\\                      
& 1.50  & 5.051 & -0.015 & 3.794 & -0.011 & 0.310\\                     
& 2.00  &  8.155 & -0.021 & 6.127 & -0.016 & 0.453 \\
\hline                     
\multirow{7}{*}{0.80} 
& 0.50 &  0.989 & -0.055 & 0.743 & -0.041 & 0.241 \\  
& 1.00  & 3.450 & -0.143 & 2.592 & -0.108 & 0.718 \\                           
& 1.50  & 6.659 & -0.205 & 5.003 & -0.154 & 1.171 \\                      
& 2.00  & 10.227 & -0.241 & 7.683 & -0.181 & 1.534 \\
\hline\hline
\end{tabular}
\end{centering}
}	
	\caption{Deviation of the lensing observables of LMRBH black holes from Kerr black hole for supermassive black holes at the center of nearby galaxies for $a=-0.40$, $a=0.40$ and $a=0.80$. Here $\Delta(X)=X_{\text{LMRBH}}-X_{\text{Kerr}}$ 
\label{table2}  
	}
\end{table*} 

\begingroup
\begin{table*}[tbh!]
	\begin{ruledtabular}
		\begin{tabular}{c c c c c c}  
				Galaxy   &           $M( M_{\odot})$      &          $D_{OL}$ (Mpc)   &     $M/D_{OL}$ & $\Delta T^s_{2,1}(\text{Kerr})$&$\Delta T^s_{2,1}(\text{LMRBH})$           \\
			\hline
		
			Milky Way& $  4.3\times 10^6	 $ & $0.0083 $ &       $2.471\times 10^{-11}$ & $14.3254 $ &  $23.2562 $     \\
             M87&$ 6.15\times 10^{9} $&$ 16.68 $
&$1.758\times 10^{-11}$& $20488.6 $ &  $33261.8$\\			
		
			 NGC 4472 &$ 2.54\times 10^{9} $&$ 16.72 $
&$7.246\times 10^{-12}$& $8461.98$ &  $13737.4$\\
			
			 NGC 1332 &$ 1.47\times 10^{9} $&$22.66  $
&$3.094\times 10^{-12}$& $4897.29$ &  $7950.38$\\
		
			 NGC 4374 &$ 9.25\times 10^{8} $&$ 18.51 $
&$2.383\times 10^{-12}$& $3081.63$ &  $5002.79$\\
			
			NGC 1399&$ 8.81\times 10^{8} $&$ 20.85 $
&$2.015\times 10^{-12}$& $2935.04$ &  $4764.82$\\
			 
			  NGC 3379 &$ 4.16\times 10^{8} $&$10.70$
&$1.854\times 10^{-12}$& $1385.9$ &  $2249.9$\\
			
			 NGC 4486B &$ 6\times 10^{8} $&$ 16.26 $
&$1.760\times 10^{-12}$ & $1998.89$ &  $3245.05$\\
		
			 NGC 1374 &$ 5.90\times 10^{8} $&$ 19.57 $ &$1.438\times 10^{-12}$& $1965.58$ &  $3190.97$\\
			    
			NGC 4649&$ 4.72\times 10^{9} $&$ 16.46 $
&$1.367\times 10^{-12}$& $15724.6$ &  $ 25527.7 $\\
		
			NGC 3608 &$  4.65\times 10^{8}  $&$ 22.75  $ &$9.750\times 10^{-13}$& $1549.14$ &  $2514.91$\\
		
			 NGC 3377 &$ 1.78\times 10^{8} $&$ 10.99$
&$7.726\times 10^{-13}$ & $593.005$ &  $962.699$\\
		
			NGC 4697 &$  2.02\times 10^{8}  $&$ 12.54  $ &$7.684\times 10^{-13}$& $672.96$ &  $1092.5$\\
			 
			 NGC 5128 &$  5.69\times 10^{7}  $& $3.62   $ &$7.498\times 10^{-13}$& $189.562$ &  $307.739$\\
			
			NGC 1316&$  1.69\times 10^{8}  $&$20.95   $ &$3.848\times 10^{-13}$& $563.021 $ &  $914.023$\\
			
			 NGC 3607 &$ 1.37\times 10^{8} $&$ 22.65  $ &$2.885\times 10^{-13}$& $456.414 $ &  $740.953$\\
			
			NGC 4473 &$  0.90\times 10^{8}  $&$ 15.25  $ &$2.815\times 10^{-13}$& $299.834$ &  $486.758$\\
			
			 NGC 4459 &$ 6.96\times 10^{7} $&$ 16.01  $ &$2.073\times 10^{-13}$ & $231.871 $ &  $376.426$\\
		
			M32 &$ 2.45\times 10^6$ &$ 0.8057 $
&$1.450\times 10^{-13}$ & $8.16214 $ &  $13.2506$    \\
			
			 NGC 4486A &$ 1.44\times 10^{7} $&$ 18.36  $ &$3.741\times 10^{-14}$ & $47.9734$ &  $77.8812$\\
			 
			NGC 4382 &$  1.30\times 10^{7}  $&$ 17.88 $  &$3.468\times 10^{-14}$& $43.3093$ &  $70.3094$\\
		
			CYGNUS A &$  2.66\times 10^{9}  $&$ 242.7 $  &$1.4174\times 10^{-15}$& $8861.76$ &  $14386.4$\\
		\end{tabular}
	\end{ruledtabular}
\caption{ Estimation of  time delay for supermassive black holes at the center of nearby  galaxies in the case  Kerr ($a=0.8$) and LMRBH ($a=0.8$ and $l=2.0$). Mass ($M$) and distance ($D_{OL}$) are given in the units of solar mass and Mpc, respectively. Time Delays are expressed in minutes.
	}\label{table3} 
\end{table*}
\endgroup
In gravitational lensing, the light is deflected while maintaining the surface brightness, but the appearance of the solid angle changes, enhancing the brightness of the images. The magnification, for the $n-$loop images, is evaluated as the quotient of the solid angles subtended by the $n^{th}$ image and the source as \cite{Bozza:2002af,Bozza:2002zj}
\begin{eqnarray}\label{mag}
\mu_n &=&  \frac{1}{\beta} \Bigg[\frac{u_m}{D_{OL}}(1+e_n) \Bigg(\frac{D_{OL}+D_{LS}}{D_{LS}}\frac{u_me_n}{D_{OL} \bar{a}}  \Bigg)\Bigg].
\end{eqnarray}
\begin{figure*}
	\begin{centering}
		\begin{tabular}{p{9cm} p{9cm}}
		    \includegraphics[scale=0.75]{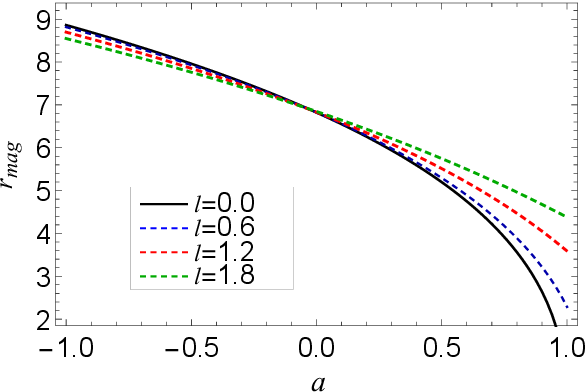}&
		    \includegraphics[scale=0.75]{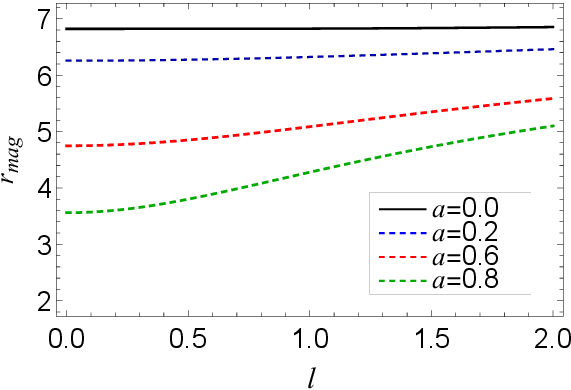}
			\end{tabular}
	\end{centering}
	\caption{Behavior of strong lensing observable $r_{\text{mag}}$, for LMRBH spacetime as a function of the parameters $a$ and $l$. It is independent of the black hole's mass or distance from the observer.}\label{plot8}
\end{figure*}
The magnification $\mu_n$ is inversely proportional to $D_{OL}^2$ and so the images are faint. But they can be bright in the limit $\beta \to 0$, i.e., when the source, lens and observer are perfectly aligned. When $\beta \to 0$, the Eq.~(\ref{mag}) diverges, suggesting that the perfect alignment maximises the possibility of the detection of the images. The brightness of the first image is dominant over the other images, as a result, we will focus on the most straightforward scenario, in which just the outermost image, $\theta_{1}$, is resolved as a single image, while the subsequent images are all crammed together at $\theta_{\infty}$. In practice, if the 1-loop image can be distinguished from the rest packed images, we can have three characteristic observables~ \cite{Bozza:2002zj} as 
\begin{eqnarray}
\theta_\infty &=& \frac{u_m}{D_{OL}},\label{theta}\\
s &=& \theta_1-\theta_\infty \approx \theta_\infty ~\text{exp}\left({\frac{\bar{b}}{\bar{a}}-\frac{2\pi}{\bar{a}}}\right),\label{sep}\\
r_{\text{mag}} &=& \frac{\mu_1}{\sum{_{n=2}^\infty}\mu_n } \approx 
\frac{5 \pi}{\bar{a}~\text{log}(10)}\label{mag1}.
\end{eqnarray} 
In the above expression, $\theta_{1}$ is the angular position of the outermost single image, $\theta_{\infty}$ is the angular position of the rest packed images, $s$ is the angular separation between the $\theta_{1}$ and $\theta_{\infty}$, $r_{\text{mag}}$ is the ratio of the flux of the first image and the all  other images. 

There is a time discrepancy between the two images because the time taken by the light routes corresponding to the various images is different. A further significant observable known as the time delay is the interval between the creation of relativistic images that take the source's varying brightness into account. Finally, the time delay $\Delta T_{2,1}$ when the images are on the same side of the lens can be tentatively calculated as follows \cite{Bozza:2003cp}
\begin{eqnarray}
\Delta T_{2,1} \approx 2\pi u_m,
\end{eqnarray}
The source must be variable to measure the time delay, which is not a stringent condition as variable stars are common in all galaxies. It could calculate the time difference between the relativistic images by implicitly assuming these fluctuations, which will manifest themselves with a temporal phase in all images.  From an observational standpoint, the accurate measurement of the time delay has a significant advantage of dimensional measurement of the system's scale, which can be used to accurately estimate the black hole's distance. 

\section{Strong gravitational lensing by supermassive black holes, Sgr A* and M87*}\label{Sec3}
Assuming that Sgr A* and M87* as characterised by LMRBH, we study the consequences of strong gravitational lensing by these supermassive black holes. By computing observables viz.,  $\theta_{\infty}$, separation $s$, and relative magnification $r_{\text{mag}}$ for different values of parameter $l$, we contrast the lensing outcomes between Kerr and LMRBH black holes. We depict our results in Fig.~\ref{plot7}, Fig.~\ref{plot7b}  and Fig.~\ref{plot8},  while Table~\ref{table1} and Table~\ref{table2} show the lensing observables and their deviation for various values of $a$, and $l$ in comparison with Schwarzschild~($a=l=0$) and Kerr black hole~($l=0$).  Our analysis of LMRBH as the lens reveals that the angular position of images for Sgr A* and M87* is consistent with the EHT-measured angular shadow diameters of Sgr A* and M87*. The results in Table \ref{table1} and \ref{table2} show that in  the case of LMRBH, the angular positions of images  are larger than their corresponding values in GR and vary slowly concerning the position of the source $\beta$. In fact, the deviation from the Kerr black hole can go up to $10.23~\mu$as and $7.68~\mu$as, respectively, for Sgr A* and M87*, at $a=0.8$  and $l = 2.00$, an effect too tiny to be observed with current telescopes. Further, the separation $s$ in the case of LMRBH for Sgr A* and M87* range between 0.008-0.376 $\mu$as and 0.006-0.282$\mu$as, respectively.  The angular separation $s$ between the first and other packed images due to the LMRBH for Sgr A* and M87* is beyond the threshold of the current EHT observation, and we may have to wait for the next generation event horizon telescope (ngEHT) for this purpose. For higher spin values, the angular separation $s$ decreases with $l$ while it increases with $l$ at lower spin levels. The relative  magnification  of the first-order images of presented in Table \ref{table1}  using Eq.~(\ref{mag}) for black holes in  GR and LMRBH. The first-order images by LMRBH  are more highly magnified than the corresponding images of black holes in GR, and the magnification increases slowly with the parameter $l$. The LMRBH images are brighter than their spherically symmetric equivalents, as indicated by the quick decline in the flux ratio of the first image to all other images with $a$ (cf. Fig.~\ref{plot8}). We have also calculated the time delay for different black holes in nearby galaxies in Table~\ref{table3}.  The time delay of the first image from that of the second image, $\Delta T_{2,1}$, for the LMRBH as Sgr A* and M87* can reach up to $23.25$~min and $554.36$~hrs, respectively, while the deviation from the Kerr black hole for Sgr A* and M87* is $8.93$~min and $212.88$~hrs, respectively.  Observing the time delay in Sgr A* is much shorter and more difficult for measurement. In the case of M87*, the time delay can reach up to a few hundred hours, sufficient for astronomical measurements, provided we have enough angular resolution separating two relativistic images. 
\begin{figure*}
	\begin{centering}
		\begin{tabular}{p{9cm} p{9cm}}
		    \includegraphics[scale=0.78]{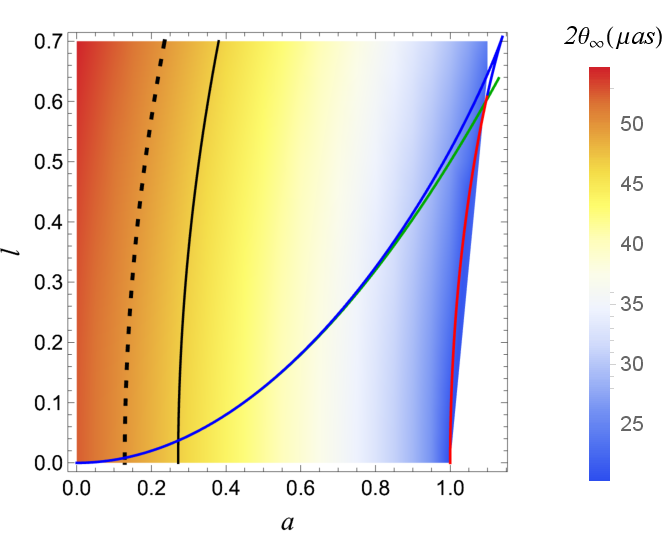}&
			\includegraphics[scale=0.78]{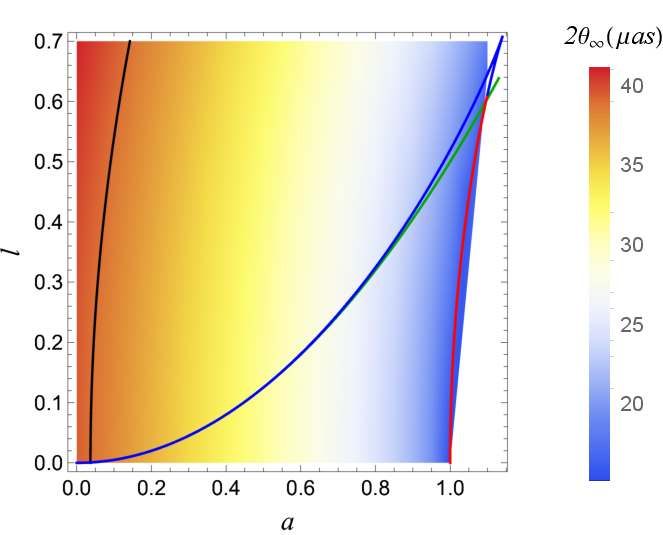}			
		\end{tabular}
	\end{centering}
	\caption{Shadow angular diameter $\theta_{sh}=2\theta_{\infty}$ of LMRBH as a function of $(a,l)$. The black and black dashed lines correspond to the Sgr A* black hole shadow at $\theta_{sh}=46.9~\mu$as and $\theta_{sh}=50~\mu$as, respectively, such that the region between these lines satisfies the Sgr A* shadow 1-$\sigma$ bound (\textit{left}). M87* shadow angular diameter when considered as a LMRBH. The black line is $\theta_{sh}=39~\mu$as, and the region contained within it meets the M87* shadow 1-$\sigma$ bound (\textit{right}).}\label{EHT}		
\end{figure*} 

\section{Constraints from EHT observations shadows of M87* and Sgr A*}\label{Sec4}
Black holes within the optically thin accreting region are anticipated to display a dark "shadow" surrounded by a bright ring \cite{bardeen1973,1973ApJ}. The boundary of the shadow is determined by the photon ring, which principally relies on the black hole's parameters \cite{Johannsen:2010ru}. The application of the shadow in understanding near-horizon geometry has sparked a flurry of activity in analyzing, both analytically and numerically, shadows for black holes in GR  \cite{Falcke:1999pj,Shen:2005cw,Yumoto:2012kz,Atamurotov:2013sca,Abdujabbarov:2015xqa,Cunha:2018acu,Kumar:2018ple,Afrin:2021ggx,Hioki:2009na, Amarilla:2010zq,Amarilla:2011fx,Amarilla:2013sj,Amir:2017slq,Singh:2017vfr,Mizuno:2018lxz,Allahyari:2019jqz,Papnoi:2014aaa,Kumar:2020hgm,Kumar:2020owy,Ghosh:2020spb,Afrin:2021wlj,Vagnozzi:2022moj,Vagnozzi:2019apd,Afrin:2021imp} 
and in LQG \cite{Afrin:2022ztr, KumarWalia:2022ddq, Islam:2022wck,Liu:2020ola,Brahma:2020eos}.  The EHT observation unveiled the shadows of supermassive black holes Sgr A* and M87* \cite{EventHorizonTelescope:2019dse,EventHorizonTelescope:2022xqj} whose sizes are within 10 percent of the Kerr predictions, furnishing another tool to explore the nature of strong-field gravity. One can put constraints on potential deviations from the Kerr, i.e., such as LMRBH stemming from LQG, as the quantum effects cannot be overlooked in the strong-field regime.  Thus, the EHT observation results of M87* and Sgr A* shadows can restrict the LMRBH-black hole parameters. We confine the parameters ($a$, $l$) within the 1-$\sigma$ level by using the apparent radius of the photon sphere ($\theta_{\infty}$) as the angular size of the black hole shadow. We model the M87* and Sgr A* as the LMRBH and use their shadow results to test the viability of our model to explain the astrophysical black holes spacetimes. 
\paragraph{Constraints from  M87* shadow}
In 2019, the EHT collaboration produced a ring of diameter $\theta_{sh}= 42\pm 3\,\mu$as as the first image of the supermassive black hole M87* \cite{EventHorizonTelescope:2019dse}. We find that, for all values of $a$ and $l$, the Kerr black hole, with a mass of $M=(6.5 \pm 0.7) \times 10^9 M_\odot$ and distance of $D_{OL}=16.8$ Mpc,  casts the smallest shadow that is contained inside the 1-$\sigma$ region \cite{EventHorizonTelescope:2019dse,EventHorizonTelescope:2019pgp,EventHorizonTelescope:2019ggy}.  Fig.~\ref{EHT} depicts the angular diameter $\theta_{sh}$ as a function of ($a,l$) for the LMRBH as M87*, with the black  corresponding to $\theta_{sh}=39~\mu$as . The LMRBH metric, when investigated with the EHT results of M87*  within the 1-$\sigma$ bound, constrains the parameters ($a, l$), viz., $0< a \le 0.03684$ and all values of $l$. Thus, based on Fig.~\ref{EHT}, LMRBH can be a candidate for the astrophysical black holes. 
\paragraph{Constraints from  Sgr A* shadow}
The EHT result for Sgr A* approximated the shadow diameter  $\theta_{sh}=(48.7\pm 7)\mu$as besides the emission ring angular diameter $\theta_d=(51.8\pm 2.3)\mu$as with the prior perceived estimates  $M=4.0^{+1.1}_{-0.6} \times 10^6 M_{o}$ and $D_{LS}=8.15\pm 0.15$ kpc \cite{EventHorizonTelescope:2022xqj}. 

The EHT observation used three independent algorithms, EHT-imaging, SIMLI, and DIFMAP, to find out that the averaged measured value of the angular shadow diameter lies within the range $\theta_{sh} \in (46.9, 50)~\mu$as. The 1-$\sigma$ interval is $\in$ $(41.7 ,55.6)~\mu$as. The average measured value of the angular shadow diameter is within the range of $\theta_{sh} \in (46.9, 50) \mu$as, according to the EHT observation, which employed three independent techniques, EHT-imaging, SIMLI, and DIFMAP. The average strongly constrains the parameters $a$ and $l$ for the LMRBH observed angular diameter of the EHT observation of Sgr A* black hole i.e., $\theta_{sh} \in (46.9, 50) \mu$as, which falls within the $1- \sigma$ confidence region with the observed angular diameter such that $0.128662 \le a \le 0.27171$ and all values of $l$ are allowed for the LMRBH. Thus, within the finite parameter space, LMRBH agrees with the EHT results of Sgr A* black hole shadow (cf. Fig.~\ref{EHT}).

\section{Conclusions}\label{Sec5}
LQG faces challenges in testing its predictions because its framework lacks rotating black hole models. We recently addressed this issue, used a nonrotating LQG black hole as a starting point, and applied the modified Newman-Janis algorithm to construct LMRBH with an additional parameter $l$ \cite{Kumar:2022vfg,Islam:2022wck}. This solution encompasses the Kerr black hole in the limit $l \to 0$. It can describe different regions of the parameter space, including multi-horizon rotating regular black holes \cite{Kumar:2022vfg,Islam:2022wck}. The spacetime singularities are artefacts of classical GR, as the prominent singularity theorems also predicted them. However, it is a widespread belief that singularities do not exist in nature and are the limitations of general relativity. Therefore, it is instructive to explore gravitational lensing by LMRBH to assess the dependence of observables on the parameter $l$  and compare the results with those for the Kerr black holes. 

With this motivation, we have analyzed the strong gravitational lensing of light because of LMRBH, which, besides the mass $M$ and angular momentum $a$, has an additional deviation parameter $l$. We have examined the effects of the LQG  parameter $l$ on the light deflection angle $\alpha_{D}(\theta)$ and lensing observables $ \theta_{\infty} $, $s$, $r_{\text{mag}}$,  $u_{m}$, in the strong field observation, because of LMRBH and compared them to the Kerr black holes.  By studying the observational implications of the gravitational lensing by this LMRBH, we can constrain the fundamental parameter $l$ of LQG. 
Indeed, we can achieve the analysis of strong fields and the testing of theories of gravity using gravitational lensing, which is a powerful tool. Comparing lensing by different gravities within the strong-field limit is valuable. In this scenario, the LMRBH model provides an exciting opportunity to discuss the observational signatures of LQG over the Kerr black hole of general relativity. Our investigation focuses on the gravitational lensing of light around the LMRBH in the strong deflection limits, which reveals that photons get deflected from their straight path and result in multiple images of a source depending on the impact parameter $u$ of the photon. At $u=u_{ps}$, photons follow circular orbits around the black hole, and the deflection angle diverges. Intrestingly,  the unstable photon orbit radius $x_{ps}$, the critical impact parameter $u_{ps}$ increase with $l$. We also found an increase in deflection angle with $l$ such that the LMRBH leads to a larger deflection angle than the Kerr black hole, and the deflection angle diverges at larger $u_{ps}$ for larger $l$ while at smaller $u_{ps}$ for larger $a$.

By considering the spacetime to be defined by the LMRBH, we estimated the lensing observables $\theta_{\infty} $, $s$ and $r_{\text{mag}}$ of the relativistic images for supermassive black holes, specifically Sgr A* and M87*. LMRBH exhibit potentially apparent distinctions from the black holes in general relativity in its predictions for gravitational lensing caused by supermassive black holes. The presence of parameter $l$ rapidly increases $\theta_{\infty} $   when compared to the Kerr black hole. We observe that  $\theta_{\infty}$ for Sgr A* $\in$ (16.40, \; 39.80) $\mu$as and its deviation from its GR counterpart can reach as much as  10.22 $\mu$as. While for M87* ranges between  12.32 - 29.91 $\mu$as, and deviation is as high as  7.68 $\mu$as. The separation $s$ decreases, with increasing $l$, for the prograde motion while it shows opposite behaviour for the retrograde motion. Moreover, the deviation in separation $s$ with $l$ for retrograde motion is much smaller than the prograde motion.  The separation $s$  due to LMRBH  black holes for Sgr A* and M87* range between 0.008-0.376 $\mu$as and 0.006-0.282 $\mu$as, respectively with the deviation from Kerr black hole can go as much as 0.241 $\mu$as for SgrA* and 0.181 $\mu$as for M87*. Compared to the analogous images of black holes in GR, the first-order images of LMRBH are significantly magnified. With increasing $l$, the magnification increases for the prograde motion but shows opposite behaviour for the retrograde motion with the relative magnitude $r_{mag}$ $\in$  (0.047,\; 1.54). Also, the deviation in magnification with $l$ for retrograde motion is smaller than the prograde motion. Finally, the time delay $\Delta T_{2,1}$ for Sgr A* and M87* deviate significantly from the Kerr black hole by $8.93$ min and $212.8$ hrs, respectively, and for the LMRBH as Sgr A* and M87*, $\Delta T_{2,1}$ can reach up to $23.25$ min and $554.3$ hrs, respectively. The time delay in Sgr A* is smaller and more challenging to measure. If we have enough angular resolution to distinguish between two relativistic images, the time delay with M87* can be as little as a few hundred hours, which is sufficient for astronomical measurements. We show how the shadow-size measurements significantly constrain the deviation parameters for the LMRBH. From M87* results of EHT, we found that  $0< a \le 0.03684  $ and all values of $l$ are allowed. The EHT results of SgrA* strictly allow $0.128662 \le a \le 0.27171$ and all values of $l$.  Thus, the LMRBH parameters $a$ and $l$ are  constrained by the EHT bounds on $\theta_{sh}$ of SgrA* and M87*, within the $1 \sigma$ region, such that a significant portion of the BH-I and BH-II parameter space and small portion of BH-III agrees with the EHT results of SgrA* and M87*,  whereas the NH spacetime is entirely ruled out.  We have found that the results gained by the Event Horizon Telescope (EHT) for Sgr A* impose stricter constraints on the parameter space of LMRBH than the limits established by the EHT results for M87*.

The primary constraint of our approach is that the LMRBH metric is not a result of the direct loop quantization of the Kerr spacetime. The LMRBH has thrilling properties, e.g., a transition surface substitutes the classical ring singularity in the Kerr black hole and provides a singularity solution of the Kerr black hole. Thus, we can expect that LMRBH captures some description of LQG. Our analysis reveal that LMRBH being BH-I with one horizon and BH-II with Cauchy and the event horizons are also observationally favoured with no restriction LQG parameter $l$. 

The results presented here generalize previous discussions on black hole lensing in GR and our results go over to Kerr and Schwarzschild black holes in the limits, $l \to 0$, and $a,\;l \to 0$ respectively. Although resolving the order estimated in SDL is challenging, the outlook for future observations looks bright. 
\section{Acknowledgments} 
S.U.I and S.G.G. are supported by SERB-DST through project No.~CRG/2021/005771. J.K. would like to thank CSIR for providing SRF.
\bibliography{LQG}
\bibliographystyle{cmpj} 
\end{document}